\documentclass[runningheads]{llncs}
\usepackage{graphicx}
\usepackage{amssymb}
\usepackage{amsmath}
\usepackage{booktabs}
\usepackage{multirow} 
\usepackage{blindtext}
\usepackage{makecell}
\usepackage{microtype}
\usepackage[colorlinks=true]{hyperref}

\usepackage{array}
\newcolumntype{C}[1]{>{\centering\arraybackslash}p{#1}}

\begin{document}
\title{Disentangling A Single MR Modality}
\titlerunning{Disentangling A Single MR Modality}
%
\author{Lianrui~Zuo\inst{1,2} \and
Yihao~Liu\inst{1} \and
Yuan~Xue\inst{1} \and
Shuo~Han\inst{3} \and 
Murat~Bilgel\inst{2} \and 
Susan~M.~Resnick\inst{2} \and 
Jerry~L.~Prince\inst{1} \and 
Aaron~Carass\inst{1}}
\authorrunning{L.~Zuo~et~al.}
%
\institute{
Department of Electrical and Computer Engineering, \\
Johns Hopkins University, Baltimore, MD~21218,~USA \\ 
\and
Laboratory of Behavioral Neuroscience, National Institute on Aging, \\
National Institute of Health, Baltimore, MD~20892,~USA \\ 
\and
Department of Biomedical Engineering, \\
Johns Hopkins School of Medicine, Baltimore, MD~21287,~USA 
}
%
\authorrunning{L. Zuo et al.}
%
%
\maketitle              
\begin{abstract}
Disentangling anatomical and contrast information from medical images has gained attention recently, demonstrating benefits for various image analysis tasks.
Current methods learn disentangled representations using either paired multi-modal images with the same underlying anatomy or auxiliary labels~(e.g., manual delineations) to provide inductive bias for disentanglement.
However, these requirements could significantly increase the time and cost in data collection and limit the applicability of these methods when such data are not available.
Moreover, these methods generally do not guarantee disentanglement.
In this paper, we present a novel framework that learns theoretically and practically superior disentanglement from \textit{single} modality magnetic resonance images. 
Moreover, we propose a new information-based metric to quantitatively evaluate disentanglement. 
Comparisons over existing disentangling methods demonstrate that the proposed method achieves superior performance in both disentanglement and cross-domain image-to-image translation tasks.

\end{abstract}
\section{Introduction}
\label{s:introduction}
The recent development of disentangled representation learning benefits various medical image analysis tasks including segmentation~\cite{liu2020variational,ouyang2021representation,duan2022cranial}, quality assessment~\cite{hays2022evaluating,shao2022evaluating}, domain adaptation~\cite{tang2021disentangled}, and image-to-image translation~(I2I)~\cite{dewey2020disentangled,zuo2021information,zuo2021unsupervised}. 
The underlying assumption of disentanglement is that a high-dimensional observation $x$ is generated by a latent variable $z$, where $z$ can be decomposed into independent factors with each factor capturing a certain type of variation of $x$, i.e., the probability density functions satisfy $p(z_1,z_2) = p(z_1)p(z_2)$ and $z = (z_1, z_2)$~\cite{locatello2020weakly}.
For medical images, it is commonly assumed that $z$ is a composition of contrast~(i.e., acquisition-related) and anatomical information of image $x$~\cite{chartsias2019disentangled,dewey2020disentangled,zuo2021unsupervised,ouyang2021representation,liu2020variational}. 
While the contrast representations capture specific information about the imaging modality, acquisition parameters, and cohort, the anatomical representations are generally assumed to be invariant to image domains\footnote{Here we assume that images acquired from the same scanner with the same acquisition parameters are from the same domain.}.
It has been shown that the disentangled domain-invariant anatomical representation is a robust input for segmentation~\cite{ouyang2021representation,chartsias2019disentangled,liu2020variational,ning2021new}, and the domain-specific contrast representation provides rich information about image acquisitions. 
Recombining the disentangled anatomical representation with the desired contrast representation also enables cross-domain I2I~\cite{zuo2021unsupervised,lyu2021dsegnet,ouyang2021representation}. 

Disentangling anatomy and contrast in medical images is a nontrivial task.
Locatello~et~al.~\cite{locatello2019challenging} showed that it is theoretically impossible to learn disentangled representations from independent and identically distributed observations without inductive bias~(e.g., domain labels or paired data). 
Accordingly, most research efforts have focused on learning disentangled representations with image pairs or auxiliary labels.
Specifically, image pairs introduce an inductive bias that the two images differ exactly in one factor of $z$ and share the remaining information. 
Multi-contrast or multi-scanner images of the same subject are the most commonly used paired data.
For example, T$_1$-weighted~(T$_1$-w)/T$_2$-weighted~(T$_2$-w) magnetic resonance~(MR) images~\cite{ouyang2021representation,dewey2020disentangled,zuo2021information}, MR/computational tomography  images~\cite{chartsias2019disentangled}, or multi-scanner images~\cite{liu2020variational} of the same subject are often used in disentangling contrast and anatomy. 
The underlying assumption is that the paired images share the same anatomy~(domain-invariant) while differing in image contrast~(domain-specific). 
The requirement of paired training images with the same anatomy is a limitation due to the extra time and cost of data collection. 
Even though such paired data are available in some applications---for example, paired T$_1$-w and T$_2$-w images are routinely acquired in MR imaging---registration error, artifacts, and difference in resolution could violate the fundamental assumption that only one factor of $z$ changes between the pairs. 
As we show in Sec.~\ref{sec:experiments}, non-ideal paired data can have negative effects in disentangling.

Labels, such as manual delineations or domain labels, usually provide explicit supervision in either the disentangled representations or synthetic images generated by the I2I algorithm for capturing desired properties. Chartsias~et~al.~\cite{chartsias2019disentangled} used manual delineations to guide the disentangled anatomical representations to be binary masks of the human heart.
In \cite{huang2018multimodal,jha2018disentangling,ning2021new}, researchers used domain labels with domain-specific image discriminators and a cycle consistency loss to encourage the synthetic images to be in the correct domain and the representations to be properly disentangled. 
Although these methods have shown encouraging performance in I2I tasks, there is still potential for improvements. The dependency of pixel-level annotations or domain labels limits the applicability, since these labels are sometimes unavailable or inaccurate.
Additionally, the cycle consistency loss is generally memory consuming and found to be an over-strict constraint of I2I~\cite{amodio2019travelgan}, which leads to limited scalability when there are many image domains.

Can we overcome the limitations of current disentangling methods and design a model that does not rely on paired multi-modal images or labels and is also scalable in a large number of image domains?
Deviating from most existing literature that heavily relies on paired multi-modal images for training, we propose a \textbf{single modality disentangling framework.}
Instead of using domain-specific image discriminators or cycle consistency losses, we design a novel distribution discriminator that is shared by all image domains for \textbf{theoretically and practically superior disentanglement~$p(z_1,z_2) = p(z_1)p(z_2)$.}
Additionally, we present \textbf{an information-based metric to quantitatively measure disentanglement.} 
We demonstrate the broad applicability of the proposed method in a multi-site brain MR image harmonization task and a cardiac MR image segmentation task.
Results show that our single-modal disentangling framework achieves performance comparable to methods which rely on multi-modal images for disentanglement. 
Furthermore, we demonstrate that the proposed framework can be incorporated into existing methods and trained with a mixture of paired and single modality data to further improve performance.

\section{Method}
\subsection{The single-modal disentangling network}
\begin{figure}[!tb]
    \centering 
    \includegraphics[width=0.65\textwidth]{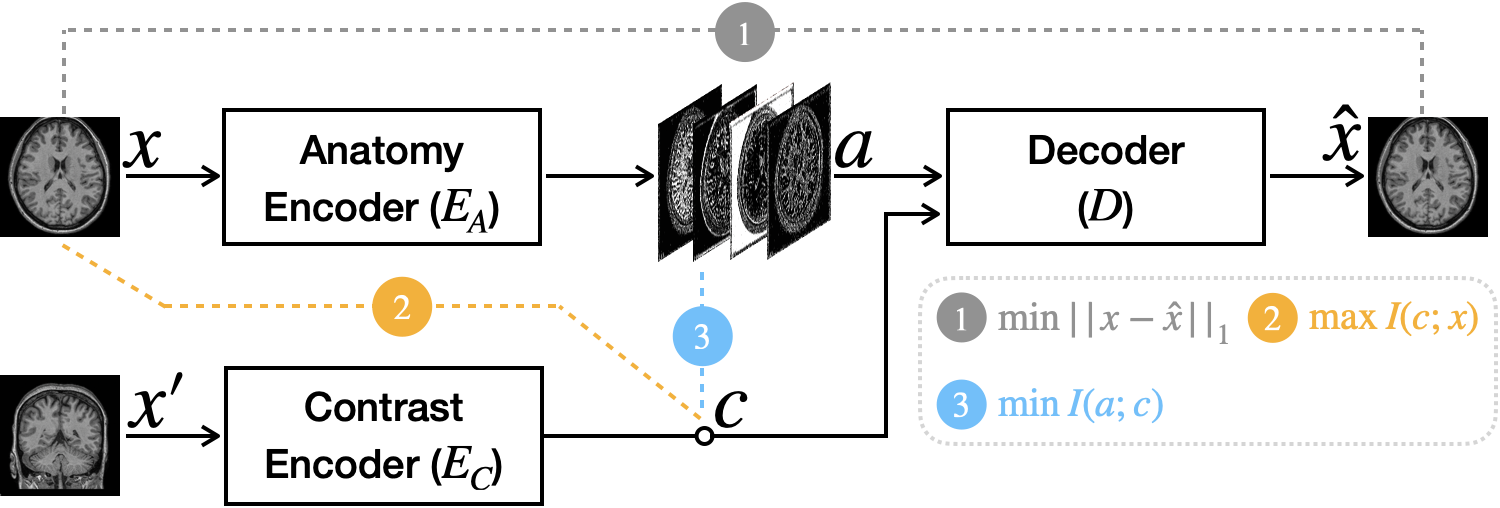}
    \caption{The proposed disentangling framework has an encoder-decoder like structure. $x$ and $x'$ are two slices from different orientations of the same $3$D volume, which we assume embed the same contrast but different anatomy information. $I(\cdot;\cdot)$ denotes MI.}
    \label{fig:framework}
\end{figure}

\textbf{General framework}
As shown in Fig.~\ref{fig:framework}, the proposed method has an autoencoder-like structure, where the inputs $x$ and $x'$ are two slices from different orientations of the same $3$D volume. $I(\cdot;\cdot)$ denotes mutual information~(MI). Note that the proposed method can be applied to datasets of other organs, as shown in Sec.~\ref{sec:experiments}; we use brain MR images for demonstration purposes. The anatomy encoder has a U-Net~\cite{ronneberger2015u} structure similar to \cite{chartsias2019disentangled,dewey2020disentangled,liu2020variational}. It generates one-hot encoded anatomical representations $a \in \mathbb{R}^{H\times W \times M}$, where $H$ and $W$ are with the same spatial dimension of $x$, and $M$ is the number of channels. The contrast encoder 
is composed of a sequence of convolutional blocks with output contrast representations $c \in \mathbb{R}^2$. $c$ is then broadcast to $H\times W \times 2$ and concatenated with $a$ as the input to the decoder, which is also a U-Net. 
The same networks are shared by all image domains, so the model size stays constant.

The overall objectives of the framework are 1) to disentangle anatomical and contrast information from input images without multi-modal paired data or labels and 2) to generate high quality synthetic images based on the disentangled representations.
The above two objectives can be mathematically summarized into three terms~(also shown in Fig.~\ref{fig:framework}) that we show are optimizable by loss functions, i.e.,
\begin{equation}
    \label{eq:objective}
    \mathcal{L} \triangleq \lambda_1 ||x - \hat{x}||_1 - \lambda_2 I(c;x) + \lambda_3 I(a;c),
\end{equation}
where $\lambda$'s are the hyperparameters and $||x - \hat{x}||_1$ is the $l_1$ reconstruction loss that encourages the generated image $\hat{x}$ to be similar to the original image $x$. 
The second term $-I(c;x)$ encourages $c$ to capture as much information of $x$ as possible. 
Since $c$ is calculated from $x'$ instead of $x$ with $E_C(\cdot)$ and the shared information between $x'$ and $x$ is the contrast, maximizing $I(c;x)$ guides $E_C(\cdot)$ to extract contrast information.
Lastly, minimizing $I(a;c)$ penalizes $a$ and $c$ from capturing common information, thus encouraging disentanglement. 
Since $c$ captures contrast information by maximizing $I(c;x)$, this helps the anatomy encoder $E_A(\cdot)$  learn anatomical information from the input $x$. 
This can also prevent a trivial solution where $E_A(\cdot)$ and $D(\cdot)$ learn an identity transformation. 
In the next section, we show how the two MI terms are optimized in training. 
We also theoretically show that $a$ and $c$ are perfectly disentangled, i.e., $I(a;c)=0$, at the global minimum of $E_A(\cdot)$.

\textbf{Maximizing $\boldsymbol{I(c;x)}$} We adopt DeepInfomax~\cite{hjelm2018learning} to maximize the lower bound of $I(c;x)$ during training. This approach uses this inequality:
\begin{equation}
    \label{eq:deep_infomax}
    I(c;x) \geq \hat{I}(c;x) \triangleq \mathbb{E}_{p(c,x)} \left[ -\text{sp} \left(-T(c,x) \right) \right] - \mathbb{E}_{p(c)p(x)} \left[ \text{sp} \left( T(c,x) \right)  \right],
\end{equation}
where $\text{sp}(r) = \log(1+e^r)$ is the softplus function and $T(\cdot, \cdot)$ is a trainable auxiliary network $T(c, x) \colon \mathcal{C} \times \mathcal{X} \rightarrow \mathbb{R}$. The gradient calculated by maximizing Eq.~(\ref{eq:deep_infomax}) is applied to both $T(\cdot,\cdot)$ and $E_C(\cdot)$. Density functions $p(c,x)$ and $p(c)p(x)$ are sampled by selecting matched pairs $\{(c^{(i)},x^{(i)}) \}_{i=1}^N$ and shuffled pairs $\{(c^{(i)},x^{(l_i)}) \}_{i=1}^N$ from a mini-batch with $N$ being the batch size and $c=E_C(x')$. Note that paired multi-modal images are not required; $x$ and $x'$ are 2D slices from the same volume with different orientations. $i$ is the instance index of a mini-batch and $\{l_i\}_{i=1}^N$ is a permutation of sequence $\{1,\dots,N\}$. 

\textbf{Minimizing $\boldsymbol{I(a;c)}$} Since Eq.~(\ref{eq:deep_infomax}) provides a lower bound of MI, it cannot be used to minimize $I(a;c)$. We propose a novel way to \textit{minimize} $I(a;c)$. Inspired by the distribution matching property of generative adversarial networks~(GANs)~\cite{goodfellow2014generative}, we introduce a distribution discriminator~$U(\cdot,\cdot) \colon \mathcal{A} \times \mathcal{C} \rightarrow \mathbb{R}$ that distinguishes whether the inputs $(a,c)$ are sampled from the joint distribution $p(a,c)$ or product of the marginals $p(a)p(c)$. Note that $c$ is detached from the computational graph while minimizing $I(a;c)$, so the GAN loss only affects $U(a,c)$ and $E_A(x)$, where $E_A(x)$ tries to ``fool'' $U(a,c)$ by generating anatomical representations $a$, such that $p(a,c)$ and $p(a)p(c)$ are sufficiently indistinguishable. 

\begin{theorem}
    \label{theory:disentangle}
    $E_A(x)$ achieves global minimum $\iff$ $p(a,c) = p(a)p(c)$. 
\end{theorem}
Theorem~\ref{theory:disentangle} says $a$ and $c$ are disentangled at the global minimum of $E_A(x)$. The minmax training objective between $E_A(x)$ and $U(a, c)$ is given by
\begin{equation}
    \min_{E_A} \max_{U} \mathbb{E}_{p(a,c)}\left[ \log U(a,c) \right] + 
        \mathbb{E}_{p(a)p(c)} \left[ \log \left( 1 - U(a,c) \right) \right],
\end{equation}
where $a = E_A(x)$. Density functions $p(a,c)$ and $p(a)p(c)$ are sampled by randomly selecting matched pairs $\{(a^{(i)},c^{(i)}) \}_{i=1}^N$ and shuffled pairs $\{(a^{(i)},c^{(l_i)}) \}_{i=1}^N$, respectively.

\textbf{Implementation details} There are in total five networks shared by all domains. $E_A(\cdot)$ is a three-level~(downsampling layers) U-Net with all convolutional layers being a kernel size of $3\times 3$ convolution followed by instance normalization and LeakyReLU. $D(\cdot)$ has a similar U-Net structure as $E_A(\cdot)$ with four levels. $E_C(\cdot)$ is a five-layer CNN with $4\times4$ convolutions with stride $2$. The kernel size of the last layer equals the spatial dimension of the features, making the output variable $c$ a two-channel feature with $H = W = 1$. Both $T(\cdot,\cdot)$ and $U(\cdot,\cdot)$ are five-layer CNNs. We use the Adam optimizer in all our experiments, where our model consumed approximately $20$GB GPU memory for training with batch size $16$ and image dimension $288 \times 288$. Our learning rate is $10^{-4}$. $\lambda_1=1.0$ and $\lambda_2=\lambda_3=0.1$.

\subsection{A new metric to evaluate disentanglement}
Since MI between two perfectly disentangled variables is zero, intuition would have us directly estimate MI between two latent variables to evaluate disentanglement. 
MINE~\cite{belghazi2018mine} provides an efficient way to estimate MI using a neural network.
However, simply measuring MI is less informative since MI is not upper bounded; e.g., how much worse is $I(a;c) = 10$ compared with $I(a;c)=0.1$? 
Inspired by the fact that $I(a;c) = H(c) - H(c|a)$, where $H(\cdot)$ is entropy, we define a bounded ratio $R_I(a;c) \triangleq I(a;c) / H(c) \in [0,1]$ to evaluate disentanglement.
$R_I(a,c)$ has a nice theoretical interpretation: the \textit{proportion} of information that $c$ shares with $a$. 
Different from MIG~\cite{chen2018isolating}, which requires the ground truth factor of variations, the ratio $R_I(a;c)$ directly estimates how well the two latent variables are disentangled.

When the distribution $p(c)$ is known, $H(c)$ can be directly calculated using the definition $H(c) = -\sum p(c) \log p(c)$. To estimate $H(c)$ when $p(c)$ is an arbitrary distribution, we follow~\cite{chan2019neural}. Accordingly, $R_I(a;c)$ for unknown $p(c)$ is given by
\begin{align}
    R_I(a;c) &\triangleq \frac{I(a;c)}{H(c)} 
        = \frac{\mathcal{D}_{\text{KL}} \left[ p(a,c) || p(a) p(c) \right] }{-\mathbb{E}_{p(c)} \left[ \log q(c) \right] - \mathcal{D}_{\text{KL}} \left[ p(c) || q(z) \right] },
\end{align}
where $z \sim q(z) = \mathcal{N}(0,\mathbb{I})$ is an auxiliary variable with the same dimension as $c$. The two Kullback-Leibler divergence terms can be estimated using MINE~\cite{belghazi2018mine}. The cross-entropy term is approximated by the empirical mean $-\frac{1}{N}\sum_{i=1}^N \log q(c_i)$.

\section{Experiments and Results}
\label{sec:experiments}
We evaluate the proposed single-modal disentangling framework on two different tasks: harmonizing multi-site brain MR images and domain adaptation~(DA) for cardiac image segmentation.
In the brain MR harmonization experiment, we also quantitatively evaluate disentanglement of different comparison methods.
\begin{figure}[!tb]
    \centering
    \includegraphics[width=0.85\textwidth]{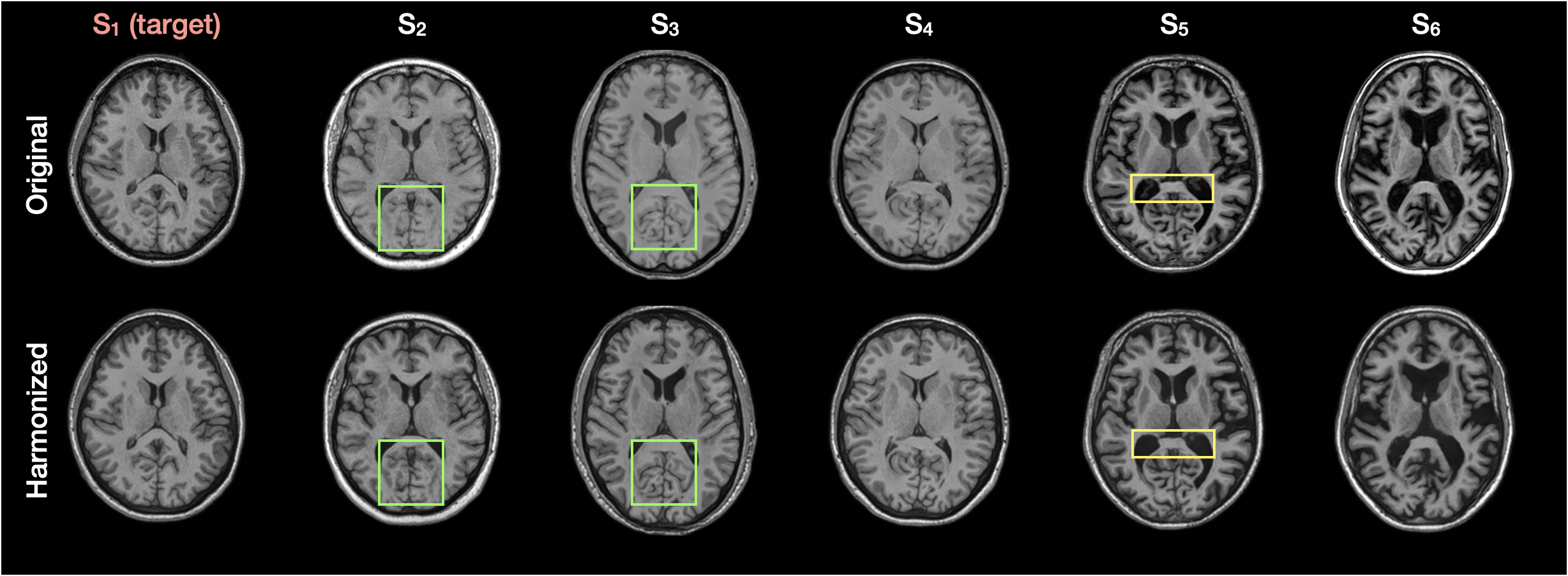}
    \caption{T$_1$-w brain MR images from ten sites~($S_1$ to $S_{10}$) are harmonized to $S_1$ using the proposed method. Six representative sites are shown due to the page limit. Green boxes highlight gray and white matter contrasts becoming more similar to the target after harmonization. Yellow boxes indicate harmonization error in the ventricles from the proposed method.}
    \label{fig:harmonized_images}
\end{figure}

\textbf{Brain MR harmonization} T$_1$-w and T$_2$-w MR images of the human brain collected at ten different sites~(denoted as $S_1$ to $S_{10}$) are used in our harmonization task.
We use the datasets~\cite{ixi,lamontagne2019oasis,resnick2000one} and preprocessing reported in~\cite{zuo2021unsupervised} after communication with the authors.
For each site, 10 and 5 subjects were used for training and validation, respectively.
As shown in Fig.~\ref{fig:harmonized_images}, the original T$_1$-w images have different contrasts due to their acquisition parameters.
We seek a T$_1$-w harmonization such that the image contrast of the source site matches the target site while maintaining the underlying anatomy. 
We have a set of held-out subjects~($N=10$) who traveled between $S_1$ and $S_2$ to evaluate harmonization.

\textit{Compare with unpaired I2I method without disentangling.}
We first compared our method trained only on T$_1$-w images (100\%\texttt{U}) with CycleGAN~\cite{zhu2017unpaired}, which conducts unpaired I2I based on image discriminators and a cycle consistency loss without disentanglement.
Results in Table~\ref{tab:compare_harmonization} show that our (100\%\texttt{U}) outperforms CycleGAN with statistical significance~($p<0.01$) in a paired Wilcoxon test. 

\textit{Does single-modal disentangling perform as well as multi-modal disentangling?} We then compared our method with two different disentangling methods which use paired T$_1$-w and  T$_2$-w for training. 
Specifically, Adeli~et~al.~\cite{adeli2021representation} learns latent representations that are mean independent of a protected variable. 
In our application, $a$ and $c$ are the latent and protected variable, respectively.
Zuo~et~al.~\cite{zuo2021unsupervised} tackles the harmonization problem with disentangled representations without explicitly minimizing $I(a;c)$.
Paired Wilcoxon tests show that our~(100\%\texttt{U}) has comparable performance with Adeli~et~al.~\cite{adeli2021representation} and Zuo~et~al.~\cite{zuo2021unsupervised}, both of which rely on paired T$_1$-w and T$_2$-w images for training (see Table~\ref{tab:compare_harmonization}).

\textit{Are paired data helpful to our method?}
We present three ablations of our method: training with all T$_1$-w images~(\texttt{$100\%$U}), training with $50\%$ paired and $50\%$ unpaired images~(\texttt{$50\%$P, $50\%$U}), and training with $100\%$ paired T$_1$-w and T$_2$-w images~(\texttt{$100\%$P}). 
The existence of paired T$_1$-w and T$_2$-w images of the same anatomy provides an extra constraint: $a$ of T$_1$-w and T$_2$-w should be identical. 
We observe in Tabel~\ref{tab:compare_harmonization} that introducing a small amount of paired multi-modal images in training can boost the performance of our method, as our (50\%\texttt{U}, 50\%\texttt{P}) achieves the best performance.
Yet, training the proposed method using all paired images has no benefit to harmonization, which is a surprising result that we discuss in Sec.~\ref{sec:discussion}.
\begin{table}[!tb]
    \centering
    \caption{Numerical comparisons between the proposed approach and existing I2I methods in a T$_1$-w MR harmonization task. SSIM and PSNR are calculated in 3D by stacking 2D axial slices and are reported as ``mean $\pm$ standard deviation''. Bold numbers indicate the best mean performance. \texttt{U}: training with unpaired data~(T$_1$-w only). \texttt{P}: training with paired T$_1$-w and T$_2$-w images.}
    \resizebox{0.9\columnwidth}{!}{
    \begin{tabular}{p{0.15\textwidth} C{0.17\textwidth} C{0.17\textwidth} C{0.17\textwidth} C{0.17\textwidth} C{0.17\textwidth} C{0.15\textwidth} }
    \toprule
        & \multirow{2}{*}{\textbf{\thead{Training\\data}}} & \multicolumn{2}{c}{\textbf{I2I: $\boldsymbol{S_1}$ to $\boldsymbol{S_2}$}} & \multicolumn{2}{c}{\textbf{I2I: $\boldsymbol{S_2}$ to $\boldsymbol{S_1}$}} & \textbf{Disentangle} \\
        \cmidrule(lr){3-4} \cmidrule(lr){5-6} \cmidrule(lr){7-7}
        &  & SSIM~($\%$) & PSNR~(dB) & SSIM~($\%$) & PSNR~(dB) & $R_I(a;c)~(\%)$ \\
        \midrule
        \texttt{Before I2I} & -- & $87.54\pm1.18$ & $26.68\pm0.77$ & $87.54\pm1.18$ & $26.68\pm0.77$ & -- \\
        \texttt{CycleGAN}~\cite{zhu2017unpaired} & \texttt{$100\%$U} & $89.62\pm1.14$ & $27.35\pm0.52$ & $90.23\pm1.12$ & $28.15\pm0.59$ & --\\
        \texttt{Adeli}~\cite{adeli2021representation} & \texttt{$100\%$P} & $89.92\pm 0.98$ & $27.47\pm 0.52$ & $90.36\pm 1.05$ & $28.41\pm 0.49$ & $5.9$ \\
        \texttt{Zuo}~\cite{zuo2021unsupervised} & \texttt{$100\%$P} & $90.63\pm 1.08$ & $27.60\pm 0.61$ & $90.89\pm 1.01$ & $28.14\pm 0.54$  & $10.8$ \\ 
        \cmidrule{1-7}
        \texttt{Ours} & \texttt{$100\%$U} & $90.25\pm 1.02$ & $27.86\pm 0.59$ & $90.52\pm 1.03$ & $28.23\pm 0.45$ & $\boldsymbol{0.1}$  \\
        \texttt{Ours} & \texttt{$50\%$U, $50\%$P}& $\boldsymbol{90.96\pm 1.00}$ & $\boldsymbol{28.55\pm 0.61}$ & $\boldsymbol{91.16\pm 1.31}$ & $\boldsymbol{28.60\pm 0.49}$ & $1.5$  \\
        \texttt{Ours} & \texttt{$100\%$P} & $90.27\pm 0.95$ & $27.88\pm 0.54$ & $90.70\pm 1.02$ & $28.59\pm 0.52$ & $3.1$ \\
    \bottomrule
    \end{tabular}}
    \label{tab:compare_harmonization}
\end{table}

\textit{Do we learn better disentanglement?} We calculated the proposed $R_I(a;c)$ to evaluate all the comparison methods that learn disentangled representations. 
Results are shown in the last column of Table~\ref{tab:compare_harmonization}. 
All three ablations of the proposed method achieve superior disentangling performance than the other methods.
Out of the five comparisons, Zuo~et~al.~\cite{zuo2021unsupervised} has the worst disentanglement between $a$ and $c$; this is likely because Zuo~et~al. only encourages disentangling between $a$ and domain labels. 
Surprisingly, $a$ and $c$ become more entangled as we introduce more T$_2$-w images in training.
A possible reason could be that T$_1$-w and T$_2$-w images carry slightly different observable anatomical information, making it impossible to completely disentangle anatomy and contrast because several factors of $z$ are changing simultaneously.
A similar effect is also reported in~\cite{trauble2021disentangled}, where non-ideal paired data are used in disentangling.

\textbf{Cardiac MR image segmentation}
\begin{figure}[!tb]
    \centering
    \includegraphics[width=0.8\textwidth]{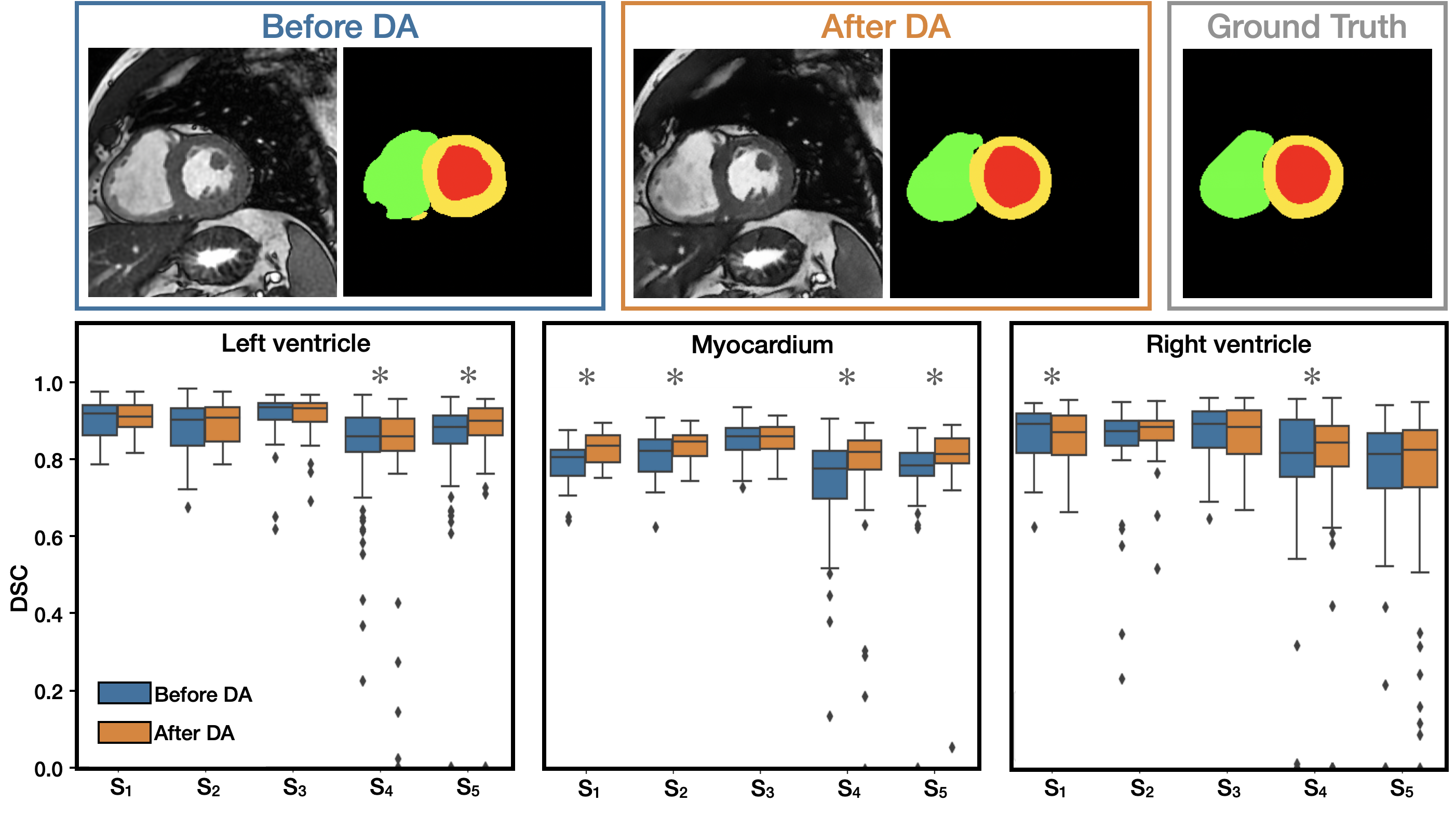}  
    \caption{Top: An example of improved segmentation from $S_4$ after DA. Bottom: DSC of multi-site cardiac image segmentation. The segmentation model was trained on $S_1$ to $S_3$, and then applied to a held-out dataset with images from all five sites. DA was conducted to translate all images to $S_3$ using the proposed method. Asterisks indicate statistically significant tests.}
    \label{fig:cardiac_segmentation}
\end{figure}
To further evaluate the proposed method, we used data from the M\&M cardiac segmentation challenge~\cite{campello2021multi}, where cine MR images of the human heart were collected by six sites, out of which five~($S_1$ to $S_5$) are available to us. 
The task is to segment the left and right ventricle and the left ventricular myocardium of the human heart.
We followed the challenge guidelines to split data so that the training data (MR images and manual delineations) only include $S_1$, $S_2$, and $S_3$, and the validation and testing data include all five sites. 
In this way there is a domain shift between the training sites ($S_1$ thru $S_3$) and testing sites ($S_4$ and $S_5$).
Since images from all five sites are available to challenge participants, DA can be applied.
Due to the absence of paired data, we only applied the proposed 100\%\texttt{U} to this task (other evaluated disentangling methods in the brain MR harmonization task cannot be applied here).

\textit{Does our method alleviate domain shift in downstream segmentation?} We adopted a 2D U-Net structure similar to the $B_1$ model reported in~\cite{campello2021multi} as our segmentation baseline.
Without DA, our baseline method achieved a performance in Dice similarity coefficient (DSC) within the top 5.
Due to domain shift, the baseline method has a decreased performance on $S_4$ and $S_5$ (see Fig.~\ref{fig:cardiac_segmentation}).
We applied the proposed method trained on MR images from all five available sites to translate testing MR images to site $S_3$ (as the baseline segmentation model has the best overall performance on the original $S_3$ images). 
Due to the poor through-plane resolution of cine MR images, we chose $x$ and $x'$ by selecting slices from two different cine time frames.
Segmentation performance was re-evaluated using images after DA.
Paired Wilcoxon test on each label of each site shows that the segmentation model has significantly improved~($p<0.01$) DSC for all labels of $S_4$ and $S_5$ after DA, except for the right ventricle of $S_5$.
Although segmentation is improved after DA, the ability of our method to alleviate domain shift is not unlimited; the segmentation performance on $S_4$ and $S_5$ is still worse than the training sites.
We regard this as a limitation for future improvement.

\section{Discussion and Conclusion}
\label{sec:discussion}
In this paper, we present a single-modal MR disentangling framework with theoretical guarantees and an information-based metric for evaluating disentanglement. 
We showcase the broad applicability of our method in brain MR harmonization and DA for cardiac image segmentation.
We show in the harmonization task that satisfactory performance can be achieved without paired data.
With limited paired data for training, our method demonstrates superior performance over existing methods. 
However, with all paired data for training, we observed decreased performance in both harmonization and disentanglement. 
We view this seemingly surprising observation as a result of the disentanglement-reconstruction trade-off reported in~\cite{trauble2021disentangled}.
This is also a reminder for future research: using paired multi-modal images for disentangling may have negative effects when the paired data are non-ideal.
Our cardiac segmentation experiment shows that domain shift can be reduced with the proposed method; all six labels improved, with five of the six labels being statistically significant. 

\subsubsection{Acknowledgement}
The authors thank BLSA participants. This work was supported in part by the Intramural Research Program of the NIH, National Institute on Aging, in part by the NINDS grants R01-NS082347 (PI: P.~A. Calabresi) and U01-NS111678 (PI: P.~A. Calabresi), and in part by the TREAT-MS study funded by the Patient-Centered Outcomes Research Institute~(PCORI/MS-1610-37115).

\bibliographystyle{splncs04}
\bibliography{cas-refs}
\end{document}